\documentstyle[preprint,aps,12pt]{revtex}
\tightenlines
\title{Bias Voltage and Temperature Dependence of Hot Electron Magnetotransport}
\author{Jisang Hong}
\address{Max-Planck-Institut f{\"u}r Mikrostrukturphysik, Weinberg 2, D-06120 Halle, Germany}
\begin{document}
\maketitle
\begin{abstract}
We present a qualitative model study of energy and temperature dependence of hot electron  magnetotransport. This model calculations are based on a simple argument that the inelastic scattering strength of hot electrons is strongly spin and energy dependent in the ferromagnets. Since there is no clear experimental data to compare with this model calculations, we are not able to extract clear physics from this model calculations. However, interestingly this calculations display that the magnetocurrent increases with bias voltage showing high magnetocurrent if spin dependent imaginary part of proper self energy effect has a substantial contribution to the hot electron magnetotransport. Along with that, the hot electron magnetotransport is strongly influence by the hot electron spin polarization at finite temperatures.
\end{abstract}
\newpage
\newcommand{\eb}{\begin{eqnarray}}
\newcommand{\ee}{\end{eqnarray}}
\section{Introduction}
An introduction of {\it hot} electron magnetoelectronic device by Monsma {\it et al} \cite{Monsma} has brought great interests in the hot electron magnetotransport. Very recently, another interesting observation has been reported by Jansen {\it et al} \cite{Jansen} at finite temperatures in the hot electron device. They obtained unusual behaviors of the collector current with temperature T depending on the relative spin orientation of the ferromagnetic layers and huge magnetocurrent even at room temperature. One should take into account transport of {\it hot} electron in the discussions of such interesting phenomena. Although the hot electron magnetotransport has not been extensively explored unlike the transport of Fermi electrons, there is an example of theoretical study of hot electron magnetotransport in a spin-valve transistor \cite{hot}. In that study, a temperature dependence of hot electron magnetotransport has been explored, and the importance of hot electron spin polarization is suggested in a spin-valve transistor. 

There have been great amount of studies in the applied bias voltage dependence of magnetoresistance in a magnetic tunneling junction (MTJ). For instance, Moodera {\it et al} \cite{Moodera} measured bias and temperature dependence of junction magnetoresistance (JMR) in the MTJ. They obtained rapid decreasing JMR with applied bias voltage, which is very intrinsic property of ferromagnetic junctions. They also explained the temperature variation by the temperature dependence of surface magnetization. Unlike large volume of data in the MTJ, only few data are available in the hot electron magnetotransport study. In the issue of energy dependence of hot electron magnetotransport, it has been presented experimentally by Mizushima {\it et el} \cite{energy}. Theoretical studies to account for the experimental observation have been also presented by the authors of the Ref. \cite{energy,account,for,the}. They claim that the inelastic scattering contributes to reduce the magnetoresistance above 1.5 eV, and the elastic scattering enhances the magnetoresistance around 1 eV. In their discussion, one should note the experimental data presented in the Ref. \cite{energy}. The Fig. 5 in the Ref. \cite{energy} shows the hysteresis curve of the sample. One can easily understand that the switching of the ferromagnetic layers is not well defined. If the switching is well defined it should occur within very narrow ranges of applied magnetic field. However, the hysteresis curve of the Fig. 5 in the Ref. \cite{energy} shows very broad features. There may be several factors contributing to broaden the hysteresis curve. For instance, the thickness of ferromagnetic layer is too thin, so that the sample may have locally different coercivity field (the thickness of Fe layers was 10 $\AA$ and 15 $\AA$ in the spin-valve base of Ref. \cite{energy}) . Therefore, it may be very difficult to extract essential physics when one explores the hot electron magnetotransport based on the data of Ref. \cite{energy}. 

Hence, in this work we shall explore the hot electron magnetotransport varying the bias voltage and temperature assuming very well defined switching of spin-valve base. Our interest is in the hot electron magnetotransport influenced by the spin dependent self energy and hot electron spin polarization in ferromagnetic materials. 

\section{Model Study}
We consider the system described in Fig. 1 to explore the issue of this work. The normal injection to the barrier surface is assumed in this model calculations. It is well known that we can write the hot electron tunneling current through the insulating barrier \cite{current} as    
\eb
I_t(eV)=\int^\infty_{-\infty}dE f_{e}(E-eV)(1-f_b(E))D_e(E)P(E)
\ee
where $f_e(E)$ and $f_b(E)$ are the Fermi-Dirac distribution functions in the emitter and base, respectively, $D_e(E)$ is the density of the states in the emitter, and $P(E)$ is the transmission probability through the barrier. It is necessary to know the exact shape of potential barrier for quantitative analysis of the tunneling current. Very recently, a ballistic electron microscopy study of aluminum barrier \cite{ballistic} for magnetic tunneling junction has been presented. It shows that the barrier height is very sensitive to the thickness of the insulating barrier. Since we have no reliable experimental data about bias and temperature dependence of hot electron magnetotransport we shall study the hot electron magnetotransport qualitative manner. We therefore take the conventional WKB approximation for $P(E)$ as 
\eb
P(E)=exp(-2\int^{w_b}_0 dx \sqrt {\frac{2m}{\hbar^2}(U_b-E-\frac{eVx}{w_b}}))
\ee 
where $U_b$ is the barrier height, and $w_b$ is the thickness of the insulating barrier. Here the energy is measured from the Fermi level of the spin-valve base. The energy of tunneled electrons are above the Fermi level of the spin-valve base, then the hot electron transport should be taken into account. The injected hot electrons will suffer from various elastic and inelastic scattering events in the first normal metal layer, however the hot electrons are not spin polarized until they reach the first ferromagnetic layer. In the ferromagnetic layer, the hot electrons have strong spin dependent self energy \cite{self}, so that the inelastic mean free path is spin dependent. Therefore, the hot electrons will be spin polarized after passing the first ferromagnetic layer. 

The issue now is the hot electron magnetotransport in ferromagnetic layer since there is no spin dependent properties in the normal metal, save for the influence on the magnitude of the collector current due to spin independent attenuation. To explore the hot electron propagation in the ferromagnetic material we need to study the Green's function $G_\sigma(\vec{k},E)$ expressed as 
\eb
G_\sigma({\vec{k},E})=\frac{1}{E-\epsilon_\sigma(\vec{k})-\Sigma_\sigma({\vec{k},E})}.
\ee
Recently, the imaginary part of spin dependent proper self energy which is related to the inelastic mean free path has been presented \cite{self}. According to the theoretical calculations, the hot electron has strong spin dependent scattering rate. This implies that the injected hot electrons (non-polarized) will be spin polarized after penetrating the first ferromagnetic layer since the attenuation is spin dependent.  By the virtue of the fact that the hot electron transport has an exponential dependence on the inelastic mean free path \cite{exponential}, the spin dependent attenuation in ferromagnets may play an essential role in the hot electron magnetotransport. To take into account the spin dependent attenuation in the ferromagnetic layer, we define the $\gamma_M(E,T)$ and $\gamma_m(E,T)$ which can be written as $\gamma_M(E,T)=exp(-w/l_M(E,T))$ and $\gamma_m(E,T)=exp(-w/l_m(E,T))$ where $w$ is the thickness of the ferromagnetic layer, and $l_{M(m)}(E,T)$ is the inelastic mean free path of majority (minority) spin electron in the ferromagnetic layer at the energy $E$ and temperature $T$. One should note that there is another Schottky barrier at the collector side, thus the energy of hot electrons should be larger than the Schottky barrier height $V_B$ to contribute to the collector current. We can then obtain the expression for the collector current  when the magnetic moments are parallel (parallel collector current) at finite temperatures
\eb
\tilde{I}^P(eV,T)&=&\int^\infty_{-\infty}dE f_{e}(E-eV)(1-f_b(E))D_e(E)P(E) \nonumber \\ 
&& \times \Gamma^3_N(E,T) \gamma_{M_1}(E,T)\gamma_{M_2}(E,T)(1+\frac{\gamma_{m_1}(E,T)}{\gamma_{M_1}(E,T)}\frac{\gamma_{m_2}(E,T)}{\gamma_{M_2}(E,T)})\Theta(E-V_B),
\ee 
and in the anti-parallel case    
\eb
\tilde{I}^{AP}(eV,T)&=&\int^\infty_{-\infty}dE f_{e}(E-eV)(1-f_b(E))D_e(E)P(E) \nonumber \\ 
&& \times \Gamma^3_N(E,T) \gamma_{M_1}(E,T)\gamma_{M_2}(E,T)(\frac{\gamma_{m_1}(E,T)}{\gamma_{M_1}(E,T)}+\frac{\gamma_{m_2}(E,T)}{\gamma_{M_2}(E,T)})\Theta(E-V_B),
\ee
where $\Gamma_N(E,T)$ accounts for the hot electron attenuation in the normal metal layer, and $\Theta$ is a step function. Since our interest in this work is to explore the energy and temperature dependence of hot electron magnetotransport due to spin dependent self energy effect, we will rewrite the Eqs. (4) and (5) in terms of hot electron spin polarization $P_H(E,T)$ \cite{hot} such as 
\eb
\tilde{I}^P(eV,T)&=&\int^\infty_{-\infty}dE f_{e}(E-eV)(1-f_b(E))D_e(E)P(E) \nonumber \\ 
&& \times \Gamma^3_N(E,T)g_1(E,T)g_2(E,T)(1+P_{H_1}(E,T))(1+P_{H_2}(E,T)) \nonumber \\
&& \times (1+\frac{1-P_{H_1}(E,T)}{1+P_{H_1}(E,T)}\frac{1-P_{H_2}(E,T)}{1+P_{H_2}(E,T)})\Theta(E-V_B),
\ee 
\eb
\tilde{I}^{AP}(eV,T)&=&\int^\infty_{-\infty}dE f_{e}(E-eV)(1-f_b(E))D_e(E)P(E) \nonumber \\ 
&& \times \Gamma^3_N(E,T)g_1(E,T)g_2(E,T)(1+P_{H_1}(E,T))(1+P_{H_2}(E,T)) \nonumber \\
&& \times (\frac{1-P_{H_1}(E,T)}{1+P_{H_1}(E,T)}+\frac{1-P_{H_2}(E,T)}{1+P_{H_2}(E,T)})\Theta(E-V_B),
\ee
where $g_i(E,T)$ is a spin averaged attenuation function in ferromagnet. In the above, the following relation has been used
\eb
\frac{\gamma_m(E,T)}{\gamma_M(E,T)}=\frac{1-P_H(E,T)}{1+P_H(E,T)}.
\ee
One can then easily obtain magnetocurrent by the definition 
\eb
MC(eV,T)=\frac{\tilde{I}^P(eV,T)-\tilde{I}^{AP}(eV,T)}{\tilde{I}^{AP}(eV,T)}
\ee

For quantitative understanding of the hot electron magnetotransport, one needs to know all the information of functions entered into Eqs. (6) and (7). Unfortunately, energy and temperature dependence of those quantities are not well known neither experimentally nor theoretically in hot electron transport we therefore shall study qualitative manner taking the values for $\Gamma_N(E,T)$ and $g_i(E,T)$ at zero temperature. We then explore the parallel and anti-parallel collector current expressed below
\eb
I^P(eV,T)&=&\int^\infty_{-\infty}dE f_{e}(E-eV)(1-f_b(E))D_e(E)P(E) \nonumber \\ 
&& \times \Gamma^3_N(E,0)g_1(E,0)g_2(E,0)(1+P_{H_1}(E,T))(1+P_{H_2}(E,T)) \nonumber \\
&& \times (1+\frac{1-P_{H_1}(E,T)}{1+P_{H_1}(E,T)}\frac{1-P_{H_2}(E,T)}{1+P_{H_2}(E,T)})\Theta(E-V_B),
\ee 
\eb
I^{AP}(eV,T)&=&\int^\infty_{-\infty}dE f_{e}(E-eV)(1-f_b(E))D_e(E)P(E) \nonumber \\ 
&& \times \Gamma^3_N(E,0)g_1(E,0)g_2(E,0)(1+P_{H_1}(E,T))(1+P_{H_2}(E,T)) \nonumber \\
&& \times (\frac{1-P_{H_1}(E,T)}{1+P_{H_1}(E,T)}+\frac{1-P_{H_2}(E,T)}{1+P_{H_2}(E,T)})\Theta(E-V_B).
\ee
The hot electron spin polarization at finite temperatures for given energy is modeled in this calculations as
\eb
P_H(E,T)=P_0(1-[T/T_c]^{3/2})
\ee  
and 
\eb
P_H(E,T)=P_0(1-[T/T_c]^2)
\ee   
where $P_0$ is the hot electron spin polarization at zero temperature, and $T_c$ is the critical temperature of the ferromagnetic material. Here, it should be pointed out that the hot electron spin polarization will be different if the thickness of ferromagnetic layer is not the same even for the same material. One can understand this remark from the Eq. (8).  

\section{Results and Discussions}
We assume that both the ferromagnetic layers in spin-valve base schematically represented in Fig.1 are Fe, and take 45 $\AA$ and 20 $\AA$ for the thickness of first and second ferromagnetic layer respectively. 20 $\AA$ is used for the thickness of the insulating barrier, 30 $\AA$ for normal metal layer, and 2.5 eV is assumed for the barrier height relative to the Fermi level of emitter material.  Here, it is of importance to note that the attenuation of low energy electron in the normal metal is around $100 \AA$ \cite{normal}. It is several times greater than that calculated in the ferromagnets \cite{self}. We therefore believe that the attenuation in ferromagnet has a substantial role in the hot electron transport. As a result, the inelastic mean free path in normal metal layer is taken as 90 $\AA$ for the energy and temperature ranges of our interest. For the spin dependent attenuation due to spin waves, Stoner excitation, and various spin non-flip processes, we adopt the results of model calculations \cite{self}. 

The Fig. 2 displays the parallel and anti-parallel collector current with the hot electron spin polarization in Eq. (12). The circle and square represent the parallel and anti-parallel collector current at zero temperature respectively, and the triangle and star stand for the parallel and anti-parallel collector current at T=300 K. One can clearly see that the parallel collector current is decreasing with temperature T, meanwhile the anti-parallel collector current is increasing with temperature T. We can understand this feature in terms of hot electron spin polarization because the $1+P_H(E,T)$ and $1-P_H(E,T)$ show the opposite behavior with temperature T, and compete each other contributing differently to the parallel and anti-parallel collector current. The calculated temperature dependence of parallel and anti-parallel collector current is different from usual behavior of current in metals because we expect that the current will be decreasing with increasing temperature T. Indeed, this kind of interesting feature of collector current with temperature T has been observed in the spin-valve transistor \cite{Jansen}. Fig. 3 represents the magnetocurrent at zero and 300 K. Fig. 3(a) is the magnetocurrent with $P_H(E,T)=P_0(1-[T/T_c]^{3/2}$, and Fig. 3(b) is the case with $P_H(E,T)=P_0(1-[T/T_c]^2)$. The magnetocurrent is increasing with applied bias voltage in both cases, which is displaying very different feature from the conventional MTJ. Qualitatively speaking, the magnetocurrent is increasing linearly roughly up to 1.2 eV, and beyond that it is starting to deviate from the linearity and almost saturated at around 2 eV . This results from the spin dependent self energy effect calculated in Ref. \cite{self}. The ratio of spin dependent inelastic mean free path is increasing with the energy of hot electron roughly speaking up to 2 or 3 eV, and the inelastic mean free path is  rapidly decreasing, which is implying strong attenuation. Hence, the hot electron with high energy does not contribute to the collector current significantly because of strong attenuation in the ferromagnet. One can also note that the magnetocurrent strongly depends on the hot electron spin polarization at finite temperatures as it is expected.

In conclusion, we have explored the applied bias voltage and temperature dependence of hot electron magnetotransport assuming well defined switching of spin-valve base. In this model calculations we have only taken into account the spin dependent self energy effect in ferromagnet. We obtain that the parallel and anti-parallel collector current have different temperature dependence, which is resulting from the hot electron spin polarization at finite temperatures. In addition, the magnetocurrent increases with applied bias voltage, and suggests an evidence of strong sensitivity to the temperature dependence of hot electron spin polarization. We hope that this work will stimulate further related issues in theoretically or experimentally.

\newpage
\begin{figure}
\caption{A schmetic display of model explored in this work. The bias voltage is applied between the emitter and base. The hot electrons are injected into the metallic base, and collected across the Schottky barrier ( with barrier height $V_B$). }
\end{figure}
\begin{figure}
\caption{The parallel and anti-parallel collector current at zero and 300 K expressed in Eqs. (10) and (11) with $P_H(T)=P_0(1-[T/T_c]^{3/2})$. Here, $T_c$ is taken as 1200 K.}
\end{figure}
\begin{figure}
\caption{The bias voltage dependence of magnetocurrent at zero and 300 K with different hot electron spin polarizatoin. Fig. 3(a) shows the magnetocurrent with $P_H(E,T)=P_0 (1-[T_c/T]^{3/2})$, and  Fig. 3(b) is for $P_H(E,T)=P_0 (1-[T_c/T]^2)$.}
\end{figure}
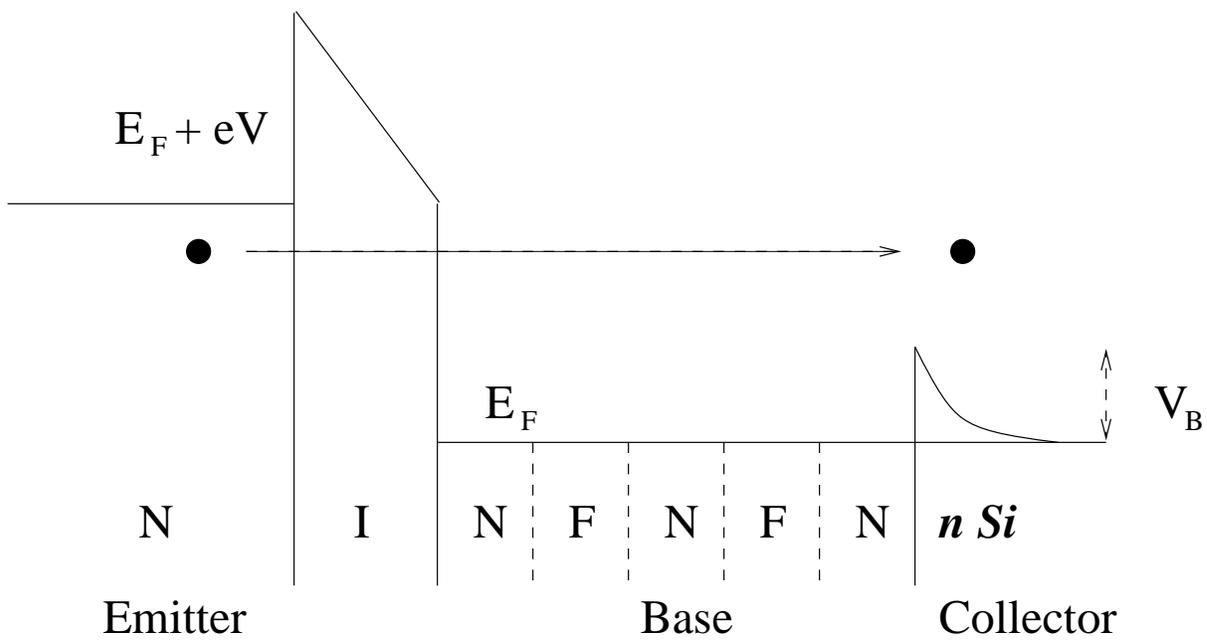
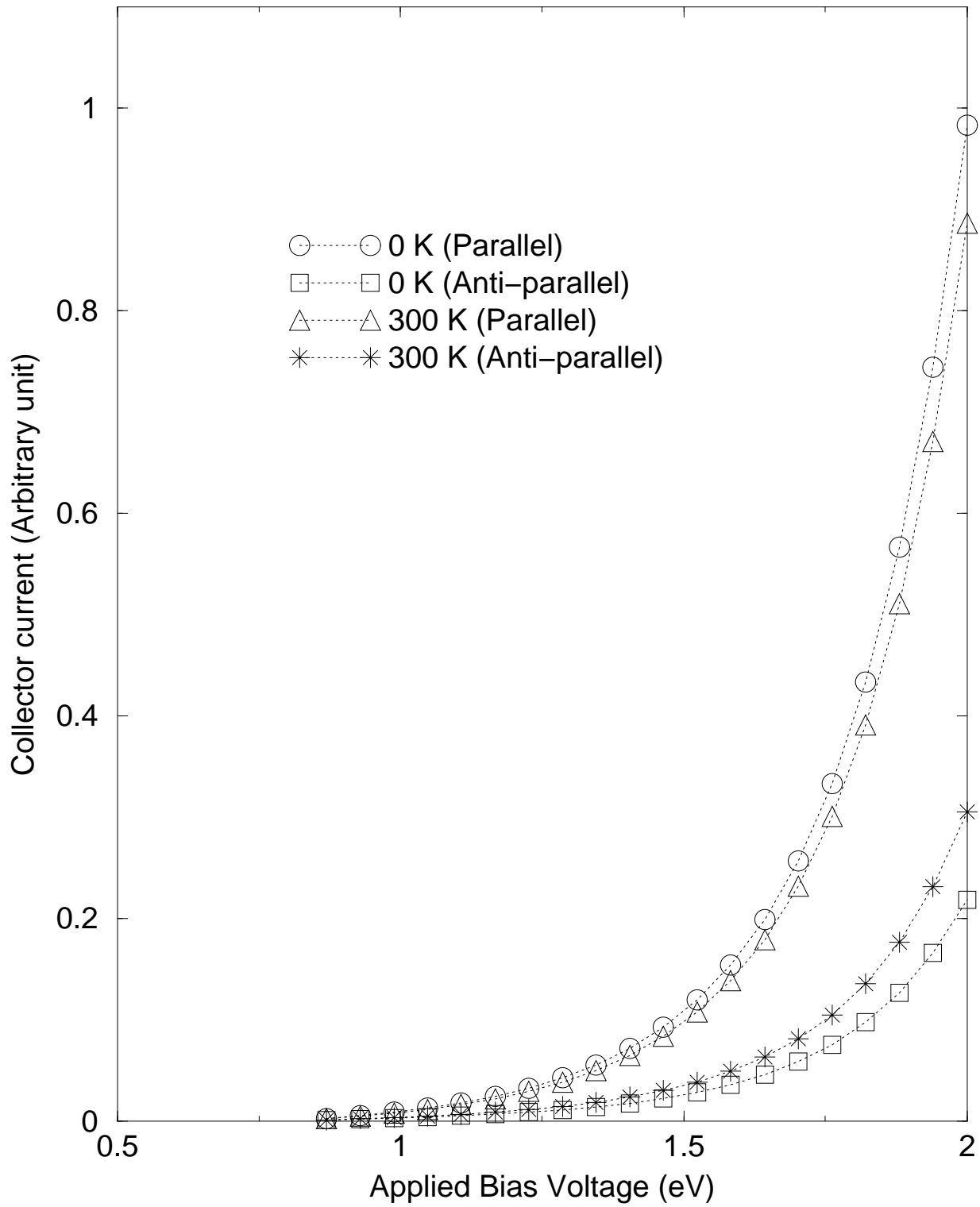
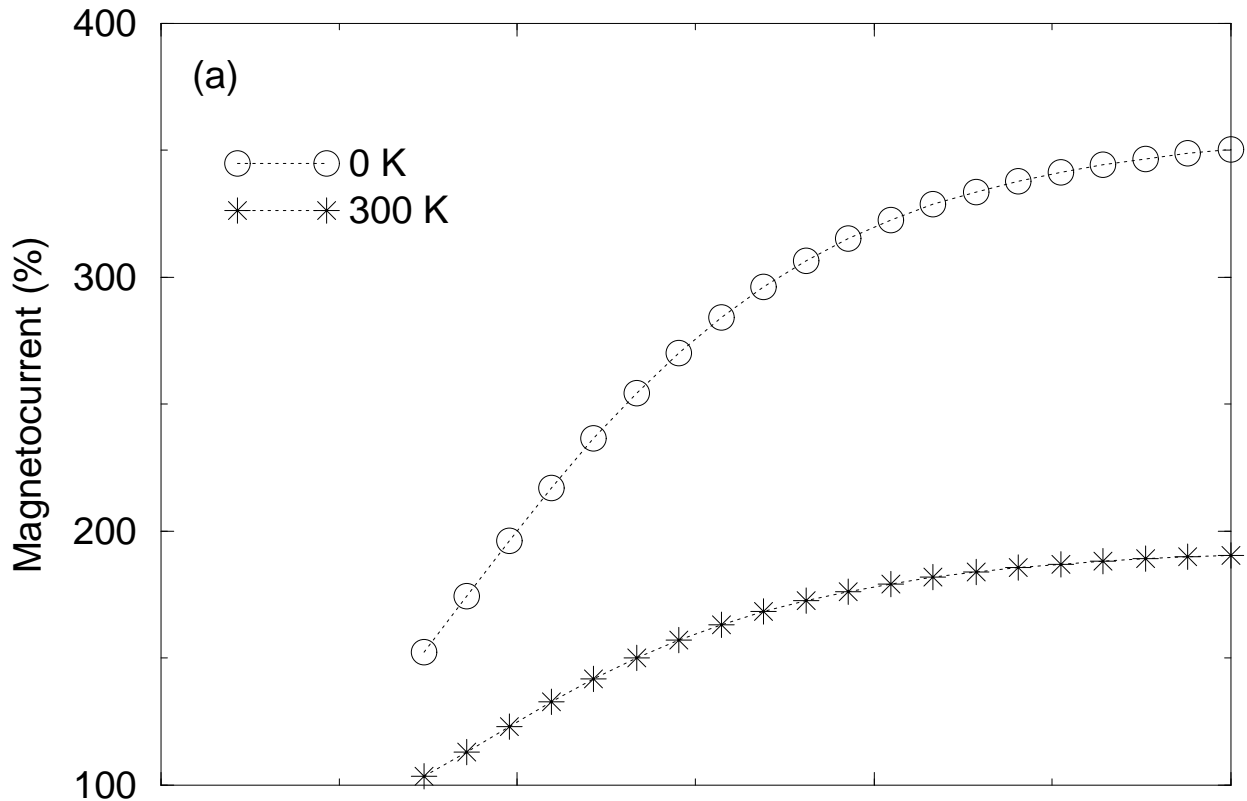
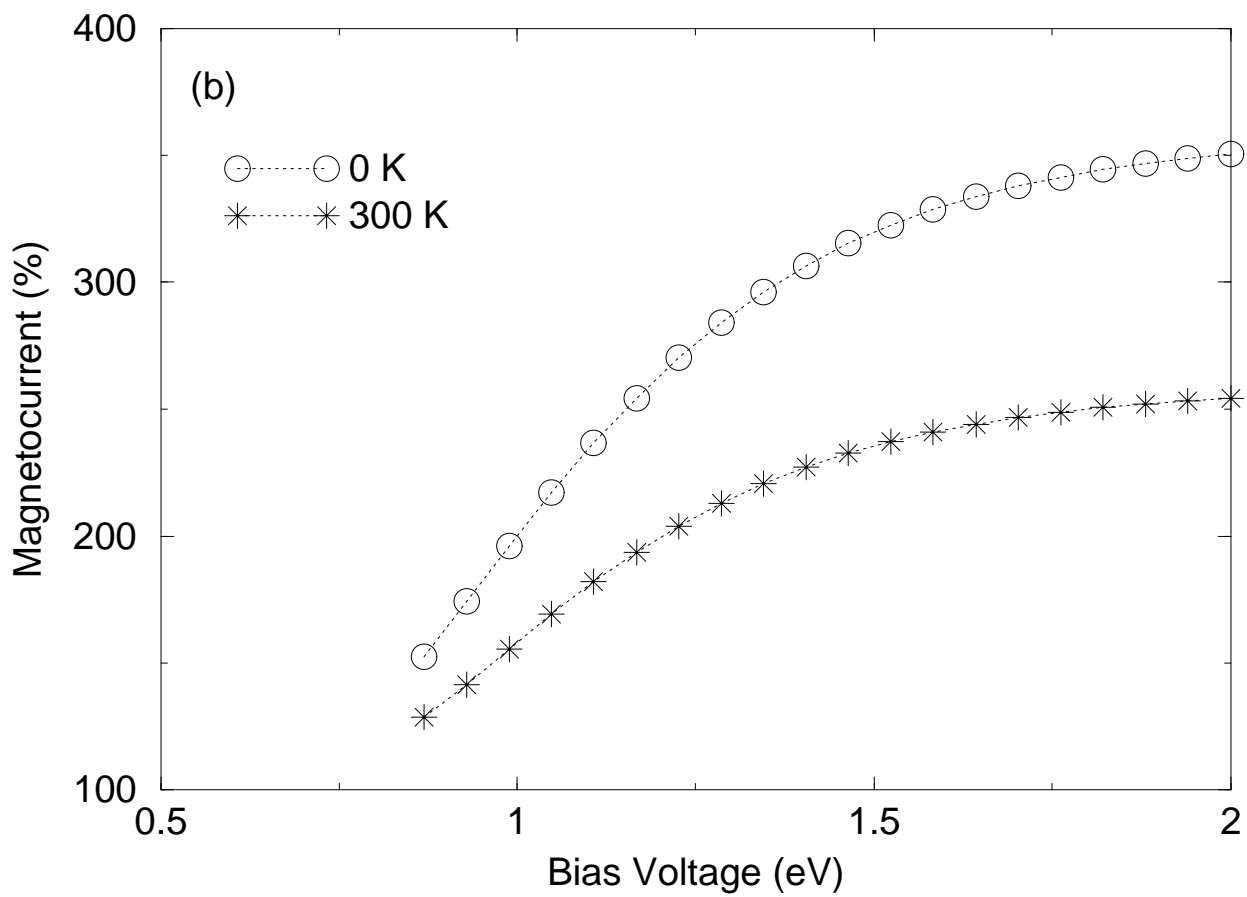
\end{document}